# Simultaneity of consciousness with physical reality: the key that unlocks the mind-matter problem

John Sanfey

Independent Researcher. London. UK

Correspondence: johnsanfey@mac.com

The problem of explaining the relationship between subjective experience and physical reality remains difficult and unresolved. In most explanations, consciousness is epiphenomenal, without causal power. The most notable exception is Integrated Information Theory (IIT), which provides a causal explanation for consciousness. However, IIT relies on an identity between subjectivity and a particular type of physical structure, namely with an information structure that has intrinsic causal power greater than the sum of its parts. Any theory that relies on a psycho-physical identity must eventually appeal to panpsychism, which undermines that theory's claim to be fundamental. IIT has recently pivoted towards a strong version of causal emergence, but macroscopic causal structures cannot be causally stronger than theirmicroscopic parts without some new physical law or governing principle. The approach taken here is designed to uncover such a principle. The decisive argument is entirely deductive from initial premises that are phenomenologically certain. If correct, the arguments prove that conscious experience is sufficient to create additional degrees of causal freedom independently of the content of experience, and in a manner that is unpredictable and unobservable by any temporally sequential means. This provides a fundamental principle about consciousness, and a conceptual bridge between it and the physics describing what is experienced. The principle makes testable predictions about brain function, with notable differences from IIT, some of which are also empirically testable.

## Section 1: Introduction

When a problem seems impossible to solve, it is worth checking its formulation. A subtle change might suggest unexpected solutions. With the problem of explaining the private, inner aspect of consciousness known as the 'hard problem' (Chalmers, 1995, 1996), I will show that insufficient attention has been paid to *time*, specifically, to the simultaneity that exists between observer and observed while reality is being experienced. This is a more precise formulation of the problem because simultaneous causation cannot have a physical explanation within the current laws of physics. However, this formulation can be solved, and in a manner that explains how consciousness exerts causal power by a mechanism that already exists in theoretical physics.

*Metaphysics* is the study of fundamental reality by identifying principles that can neither be derived from more basic principles nor reduced to something more fundamental, and which are known as *first principles*. The goal of this paper is to uncover such a principle that governs consciousness. Consciousness is unobservable, so its relationship to physics cannot be found by empirical discovery, only by argument. If such an argument is to produce a principle with fundamental credentials, its initial premisses must be completely certain, and any inferences must be deductive. I will present two sets of arguments below, a preliminary one, and a second pair that meets these criteria. In the first I will make the basic assumption that physical reality exists, together with its principle of physical causality which is the root cause of the hard problem. For the second set of arguments however, the simultaneity arguments, there are no assumptions. The initial premises for these arguments are phenomenologically certain and all inferences are deductive. Provided the arguments are correct, the principle they reveal must also be true.

I will begin by introducing the hard problem and examining how previous approaches have tried to address it. Throughout, I will pay particular attention to Integrated Information Theory (IIT) which provides the deepest explanation for consciousness at the causal level, while still failing to resolve the hard problem. Like many other theories, IIT depends on a central identity between conscious experience and some physical substrate, a *psycho-physical identity*. In IIT, subjective experience is identical to a particular type of physical substrate, defined as one with intrinsic causal power resulting from its informational structure. The problem with any psycho-physical identit is that physical realism is just one of several possible inferences that explain the invariant nature of what we experience. Once it is proven that having this choice can itself have consequences, then it is also proven that there are degrees of causal freedom in the subjective side of the psycho-physical identity that do not *necessarily* exist in the other. This does not prove IIT to be false, only that it is not fundamental, and consequently, is incapable of resolving the hard problem, which can only be solved at a fundamental level.

To be fundamental, a governing principle of consciousness must explain how objective facts emerge from first person or subjective experience (Hales & Ericson, 2022), and also how subjective experience relates to the physics describing those facts, irrespective of whether we choose physical realism or its alternatives. In doing so, the principle should explain the purpose of consciousness and how it can be produced by non-conscious (physical) components. The principle uncovered below meets these criteria and can be summarised as: there is always an observer; the observer always has a choice, and that choice always has causal consequences. The principle will establish a solid metaphysical basis for the development of *causal emergence* in neuroscience.

The main arguments are presented in Part 2. Over the course of both arguments, the hard problem will be reformulated to focus on the apparent conflict between phenomenal simultaneity and physical causality. At first sight, the re-formulated version will appear just as hard as the current one, but in fact, is readily solvable. The final section defines the bridging principle and conceptual framework, which I call *abstract realism*, that describes how consciousness can be integrated into science without invoking panpsychism, and how it can exert causal power without disturbing the laws of physics. The section also outlines its predictions for a biological explanation of consciousness and predicts some notable differences from IIT, some of which are empirically testable.

**The hard problem:**

The term *hard problem* is an ironic reference to the idea that every phenomenon in nature is potentially explicable, except subjective consciousness (Chalmers, 1995,1996). There is an '*explanatory gap'* between the phenomenon of experiencing and the physical world (Levine, 1983). The root of the problem is the fundamental principle in science that every observable phenomenon must have a physical cause. Every change in a mental state must correspond to a physical change, a concept known as *supervenience* (Davidson, 1970). However, if causality operates at the physical level rather than its corresponding mental state, then why do mental states exist at all; the physical processes of the brain should operate equally well without someone experiencing them (Chalmers, 1995, 1996)? Of course, consciousness might be non-physical, but if so, it could not cause physical effects without violating the principle of physical causality (Kim, 1990; 1999). It is conceivable that a human clone or zombie that was physically identical to the original but not conscious, would behave the same as the conscious version (Kirk, 1974)[1]. In addition to this *conceivability* argument there are *knowledge* arguments against the possibility of any physical explanation for consciousness. The first of them was developed by Broad (1925), and in Jackson's version, a scientist (Mary) who has spent their entire life in a black and white room might acquire all possible knowledge about the colour red but would still learn something new when they eventually see something red. No amount of knowledge could explain what it is like to experience redness (Jackson, 1982), and no amount of structure and function can explain the existence of subjective experience. Consciousness seems fundamentally different from the physical world.

Before the arguments proper, I would like to briefly review those previous approaches to the hard problem that have taken seriously the issue of physical causality.

**Approaches to the hard problem**

*Emergence:*

Emergence is the idea that novel physical properties appear at certain levels of complexity that cannot be predicted from their constituent parts, but which are nevertheless determined by those parts. The liquidity of water is a common example. Emergence has long been a popular explanation for consciousness (Broad, 1925; Popper & Eccles, 1977; Sperry, 1990; Searle, 1992; Silberstein, 1998; Silberstein & Mckeever, 1999; Silberstein, 2001; Chalmers, 2006; Feinberg & Mallatt, 2020), but it fell out of favour when it became clear that being unpredictable does not provide causal power. Chalmers argued that consciousness will prove to be the only example in nature where emergent properties do have own causal power, so-called *strong emergence* (Chalmers, 2006), something that would require the discovery of new physics. However, he also described a potential loophole. If the emergent properties were non-deducible as a *matter of principle* from low-level facts, but still determined by them, they would become indistinguishable from strong emergence (Chalmers, 2006). The arguments below will deliver a principle that, among other things, meets the requirements for a Chalmers loophole of strong emergence.

Higher order (HO) theories seem to require a loophole of this sort (Armstrong, 1968; Rosenthal, 1986; Lycan, 2001; Carruthers, 2005; Genaro, 1993; Van Gulick, 2004). These theories claim that a mental state is conscious when it is the subject of a simultaneous higher order representation. However, it is no coincidence that the word *simultaneous* appears in Van Gulick's description (Van Gulick, 2014/22). By casually evoking simultaneity, HO theories create the illusion of a solution without explaining how it can exist within current physical laws.

The concept of *causal emergence* grapples with a related problem. Causal emergence is the idea that higher scale causal relationships can be stronger than underlying microscopic ones without violating the principle of supervenience (Hoel et al., 2013,). The concept developed within IIT to explain how integrated information can have causal power greater than the sum of its parts (Tononi, 2004, 2008; Hoel et al., 2013, 2016; Albantakis et al., 2023), and builds on work from outside IIT (Barrett & Seth, 2011; Seth & Barrett, 2011b; Rosas et al., 2019, 2020; Mediano et al., 2022a, 2022b). But the concept is somewhat muddled at a metaphysical level. Some claim that new cause-effect properties emerge at certain levels of complexity (Marshall et al., 2018), a claim of strong emergence that would require a new physical law or governing principle. Others make it clear that the macroscopic causal properties are purely epistemic without implying strong emergence, while still referring to 'downward causation' (Mediano et al., 2022a, 2022b). In a similar vein, causal emergence has been described as the optimal scale at which the causal structure snaps into focus for an observer (Hoel, 2021). The metaphysical confusion in the field reflects a more general 'observer problem' throughout physics, one which is intimately related to the hard problem, and can only be resolved by a governing principle of consciousness.

[1] Kirk's paper was the first to mention Zombies, but the argument has roots as far back as Descartes.

*Quantum mechanics (QM)*

Quantum mechanics generally applies at the microscopic scale, to molecules, atoms, and sub-atomic particles. Its core mystery is captured in the double-slit experiment, where a particle seems to know whether it is being aimed at a single or double slit. If the former, it behaves like a particle and if the latter, it behaves like a wave. The quantum state, described by a wavefunction, seems to entail non-local information, what Einstein called 'spooky action at a distance'. Early pioneers were hopeful that these mysteries might overcome the problem of physical determinism, but that proved difficult without substance dualism whereby the quantum wavefunction interacts with mind-stuff and collapses into the precise, localised forms of classical reality (Wigner, 1961; Stapp, 1993, 2015). In Bohm's theory, the quantum wavefunction is guided by a new kind of 'active information' with non-local quantum properties, which guides fundamental particles (Bohm, 1980, 1990; Bohm & Hiley, 1993; Hiley, 2001; Pylkkänen & Hiley, 2005). The model provides for mental causation (Pylkkänen, 2019), but the outcomes of wavefunction measurement are still based on the probabilistic distribution of properties in the wavefunction and it is difficult to see how randomness can ever be helpful for explaining conscious volition (Atmanspacher, 2004/2020).

A major problem for quantum mind theories is to explaini how quantum effects can occur in the brain at a sufficient scale to be useful. Quantum processes typically concern energy levels and time intervals many orders of magnitude smaller than would be relevant in a large mass of warm, noisy brain tissue (Tegmark, 2000). However, Penrose and Hameroff have developed a plausible biological quantum theory of consciousness known as Orch OR (Orchestrated Objective Reduction). The theory integrates Penrose's twister theory of space-time with Hameroff's work on quantum coherence in cellular microtubules (Hameroff & Penrose, 1996. Hameroff, 2006). Penrose sought to explain the apparent non-computational insights of which consciousness seems capable[2]. He argued that algorithms can only follow decision rules, whereas consciousness understands them (Penrose, 1989). In Orch OR, the wavefunction collapses when the divergence of mass probabilities becomes too great to sustain. The theory postulates that each individual collapse is a moment of consciousness, and that the symmetrical arrangement of tubulin molecules in microtubules somehow *orchestrates* these moments in a useful but unexplained manner.

Without a governing principle of consciousness, Orch OR is essentially panpsychist, but, it does go some way to explaining how quantum coherence might exist long enough to play a role in the phenomenology of consciousness. Many of its predictions are empirically testable including the wavefunction collapse itself (Penrose, 2021), although this latter prediction can only be proven in terms of gravitational collapse, not for being a moment of consciousness.

*Electromagnetic (EM) field theories*

Each neuron in the brain communicates by sending an electrical charge down a long biological cable or axon. Any moving electrical charge creates an EM field, so the brain is interlaced with a complex, three-dimensional web of EM fields. Electromagnetic (EM) field theories postulate that these fields are conscious, especially those created when groups of neurons fire synchronously, a phenomenon known to be important for consciousness. Fleeting long-range synchrony lasting 230ms in the gamma range of frequency (30-80Hz), followed by active de-synchronisation is associated with conscious perception (Rodriguez et al., 1999). In Pockett's EM theory, consciousness is an epiphenomenon associated with spatial patterns in EM fields, (Pockett, 2000, 2002, 2012). Some EM theories appeal to emergence (John, 2001, 2002; Feinberg & Mallatt, 2020). Higher order (HO) theories describe a nested hierarchy of increasingly organised spatiotemporal patterns (Finglekurts et al., 2009, 2013. Young et al., 2022). Some EM theories use panpsychism to explain the 'hard' aspect of consciousness and EM fields, the easy part (Hunt, 2011, Hunt and Schooner, 2019). In others, the quantum vacuum interacts with the brain's EM fields (Keppler & Shani, 2020, Keppler, 2022).

---

[2] Penrose was strongly influenced by Libet's famous experiments in the 1980's which showed that many conscious actions are initiated unconsciously in the brain about half a second before subjects become aware of making the decision to act (Libet et al., 1983). Penrose speculated that the problems posed by Libet's experiments might be explained by time divergence implicit in the spacetime superposition of mass prior to reduction, (Penrose, 2021).

Barrett proposed modifications to IIT to account for the continuous nature of fundamental fields, specifically EM fields, a proposal known as the Field Integrated Information Hypothesis (FIIH), which will be discussed further below (Barrett, 2014, Barrett & Mediano, 2019). McFadden's conscious electromagnetic information (CEMI) theory also offers a causal explanation of consciousness. In CEMI, consciousness is the information encoded in EM patterns associated with neuronal firing. CEMI fields are postulated to be causally active on the neurons that support them. The claim is that CEMI fields differ from non-conscious EM fields by their ability to "*generate (rather than merely transmit) thoughts as gestalt (integrated) information*" (McFadden, 2000, 2002, 2013, 2020, 2023). In CEMI, the EM field can push and pull neurons toward or away from firing to achieve the desired motor actions. There is some indirect evidence that supports both CEMI and a general role for EM fields in consciousness.

Slow periodic electrical activity in the hippocampi of rabbit brains have been shown to propagate without chemical synaptic transmission and to generate electric fields that activate neighbouring cells, a process known as *ephatic* transmission (Chiang et al., 2019). Ephatic transmission is much faster than synaptic and may play a role in modulating patterns of neuronal firing across the brain (Ruffini et al., 2020). It has also been shown that working memories correlate better with the patterns of EM fields than with the neurons creating those fields (Pinotsis & Miller, 2022). In addition, the anatomical pattern of neuronal connections correlates well with mathematics specific to harmonic oscillation (Atasoy, 2016). Further evidence for EM emerged when neuronal firing patterns within brain slices were shown to be influenced by the external application of EM fields that simulate the brain's endogenous EM fields (Frohlich and McCormick, 2010; Anastassiou et al., 2011). While much of the above evidence indicates some role for EM fields in consciousness, none of it proves that consciousness is instantiated in the EM fields themselves, rather than the systems producing them.

**Neuroscience**

Until recently, neuroscience has focussed on the neural correlates of consciousness (NCC), and generally ignored the question of why NCC require subjective experience (Wu, 2018). However, once it became clear that correlating processes were often active before or after conscious events, attention turned to theories that seek to explain the phenomenon rather than simply correlating with it (Aru et al., 2012, De Graaf et al., 2012, Koch et al., 2016).

*Global Workspace Theories:*

Global Workspace Theory (GWT) was first proposed by Baars as a computational, cognitive model, (Baars, 1988, 1997, 2002), and later modified into Global Workspace Dynamics (GWD) (Baars et al., 2013, 2019, 2021). It has a neuronal version called Global Neuronal Workspace (GNW), (Dehaene, 1987, 2003). GWT used the metaphor of a theatre where the audience is the unconscious processes of the brain and where the spotlight, corresponding to conscious attention, shines on its content, the actors on the stage. The actors broadcast their message to recruit help from the audience. Conscious

‘ignition’ occurs with the recruitment of a large network of neurons in the pre-frontal and parietal regions of the brain (Dehaene, 2011). The theories have been useful for developing neural networks and AI systems (Artificial Intelligence), but they do not claim to address the hard problem.

*Predictive Processing:*

Predictive processing is the idea that the brain models the external environment and constantly updates its models through interaction with it. The idea has deep historical roots, perhaps even with Aristotle, but more recently with Helmholtz (1866) who developed the Free Energy Principle (FEP) and the concept of ‘unconscious inference’, now called FEP-AI, (Active Inference).

FEP is a mathematical principle describing how an adaptive system at equilibrium with its environment must appear to model both the external environment and its own internal states to resist the natural tendency towards disorder (Friston et al., 2007). Free energy is the surprise to a predictive system when its models encounter perceptual data (Friston, 2010). The modelling is Bayesian, based on prior and post probabilities.

Many FEP-AI theories describe a nested hierarchy of predictive systems existing within systems and separated from each other by formal boundaries known as Markov Blankets (Pearl, 1988; Rao & Ballard,1999; Clark, 2013; Safron, 2020; Ramstead et al., 2023). The simultaneity between multiple interactions is important in FEP-AI, but this creates the same metaphysical issue that HO and causal emergence theories face and is increasingly recognised as a serious challenge for the field (Safron et al., 2022). The nested hierarchy approach can lead to a ‘homunculus’ problem where the deepest Markov Blanket becomes the viewer of other screens in a cartesian theatre (Ramstead et al., 2023). Some have tried to overcome the problem by appealing to a psycho-physical equivalence between informational free energy and thermodynamic potential energy (Parr et al., 2019), or even an identity. But Kiefer admits that this would require some ‘philosophical heavy machinery’ (Kiefer, 2020). Some sidestep the hard problem directly and focus instead on making steady progress in the field (Seth, 2016). Safron takes this view and has developed a model integrating features of GWT, FEP-AI and IIT, termed the Integrated World Modelling Theory (IWMT) of consciousness (Safron, 2020). But he recognises that the hard problem can only be resolved with a fundamental bridging principle (Safron et al., 2022). IIT on the other hand, does claim to be capable of solving the hard problem (Tononi et al., 2015).

*Integrated InformationTheory (IIT):*

IIT provides a deep, causal explanation of consciousness based on a central identity between subjective experience and an information structure with intrinsic causal power. The theory has evolved over the years, (Tononi, 2004, 2008; Oizumi et al., 2014: Tononi & Koch, 2015; Tononi et al., 2022; Albantakis et al., 2022). The most recent versions include an appeal to ‘causal emergence’ (Hoel et al., 2016; Marshall et al., 2018), with further adaptions that include quantum information (Albantakis et al., 2023).

The starting point of IIT now called the *zeroth* axiom, described in version 3.0 (Oizumi et al., 2014) as the fact that subjective experience exists for certain. IIT describes phenomenal consciousness with five axioms, from which are derived five corresponding postulates about their physical substrate. The first axiom is *intrinsic existence*, that conscious experience exists and must have causal power[3]. The postulate derived from this axiom is that there must be a physical substrate with intrinsic causal power. The other four axioms are that each experience has *composition*, *Information*, *integration*, and *exclusivity* in the sense of being definite. Each moment of experience has a unique compositional structure and is integrated into a single experience.

[3] This is an inverted version of Plato’s Eleatic principle, in which anything with power, must exist (Plato, 247). IIT claims that the reverse is also true, that if something exits, then it must have causal power (Albantakis, 2022).

The 'Intrinsicality' of subjective experience is defined as 'an experience for an experiencer' (Albantakis, 2022) and its physical substrate as an informational structure whose causal power is greater than the sum of its parts. In that sense intrinsic information is 'maximally irreducible'. The degree of consciousness can be quantified using the term *Φ (phi)* which describes the extent to which a structure defined by the five postulates is *maximally irreducible* ($\Phi^{max}$) such that its causal power cannot be reduced without losing some power. Any information system that is an integrated whole, such that the causal power of the whole is greater than the sum of its parts, has Φ greater than zero, and is conscious (Koch & Tononi, 2008; Oizumi et al., 2014; Tononi & Koch, 2015).

*IIT as a fundamental theory:*

IIT has been criticised for the practical difficulty of measuring *Φ* in complex systems (Barrett & Seth, 2011; Barrett, 2016; Tegmark, 2016; Kim et al., 2018; Kim & Lee, 2019; Barrett & Mediano, 2019; Mediano et al., 2019, 2022b), a difficulty acknowledged within IIT (Oizum, 2014). Others question whether its most specific features are testable (Mediano et al., 2022b). Some criticise its lack of attention to predictive modelling (Seth et al., 2011; Barrett & Mediano, 2019; Safron, 2020). But its credentials as a fundamental theory have also been challenged. Can its axioms be fundamental when they appeal to a personal experience of consciousness while some conscious philosophers such as Dennett (2016) and Frankish (2016) claim that consciousness is an illusion (Bayne, 2018)? Are the postulates truly *deduced* from axioms when alternative inferences are also possible? Is its central identity truly an identity when $\Phi^{max}$ is both intrinsically private and also extrinsically measurable (Mørch, 2019)?

Others point out that without specifying a level of granularity, IIT's measure of maximal quantities will tend to infinity when applied to the continuous fields of fundamental physics, and that *Φ* never be zero because real interactions in physical systems are always bi-directional (Barrett, 2014, 2016; Barrett & Mediano, 2019). Tononi and Koch conceded these points while also claiming that spacetime may not be continuous at a fundamental level (Tononi & Koch, 2016). Barrett proposed a reformulation of IIT to enable measurement of intrinsic information applicable to continuous fields, which he calls the Field Integrated Information Hypothesis (FIIH) and makes some specific technical suggestions to improve IIT (Barrett, 2014; Barrett & Mediano, 2019). IIT responded to these challenges by adapting its formalism for fundamental physical fields (Hoel et al., 2016; Marshall et al., 2018) including quantum entanglement (Albantakis, 2023b).

Regarding the hard problem, IIT contains a fatal flaw on the fundamental issue of causality. One of its architects, Christof Koch, has acknowledged that some 'new insights or 're-formulation' of science may be required before the mind-matter relationship can be resolved at its deepest level (Koch, 2022; 2:45), but neither he, nor anyone else has yet articulated the metaphysical problem at IIT's core.

*IIT's metaphysical problem:*

Panpsychism is the idea that subjective experience is a fundamental feature of reality and is considered by many to be inescapable (Seagar, 2006; Strawson, 2006). It usually entails consciousness being epiphenomenal, with no causal power, but IIT is an exception to this rule, which makes it an ideal case to study.

At first glance, IIT seems little different from Chalmers' double aspect theory of information in which experience is '*information from the inside, physics is information from the outside*' (Chalmers, 1996). The umbrella term for dual-aspect theories is *neutral monism*, where fundamental reality is neither physical nor mental but a singular thing with two aspects or perspectives such as first-person and third person (Stubenberg & Wishon, 2023). Chalmers' theory is both epiphenomenal and panpsychist because it describes 'inside' as the inner substance of matter, or Kant's unknowable noumenon. In the more recent versions of IIT on the other hand, the fundamental property of consciousness is not carried by the substance of matter but exists when complexes of elements carry *Effective Information* (EI) with intrinsic causal power. The intrinsicality of subjective experience is identical to this causal information structure. But as is the case with any psycho-physical identity, the central identity in IIT cannot solve the hard problem, for the following reason.

Consciousness is more certain than anything else (IIT's zeroth axiom). The independent existence of physical reality is simply the best explanation for invariant behaviour that we observe, but it is not *certain*. IIT defines the intrinsicality of experience as an 'experience for an experiencer' (Albantakis, 2022), which implies not only a deeper duality between observer and observed within the subjective phenomenon, but also a choice to be made about what is experienced. Physical realism is just one option, and once it is proven that having this choice comes with consequences, then IIT's core identity can no longer be considered fundamental because there would be deeper degrees of causal freedom in one side of the identity that do not *necessarily* exist in the other. It will be shown below that this choice within subjectivity does indeed have observable consequences.

A fundamental theory of consciousness must account for all causal freedom at the subjective level, irrespective of which inference we choose concerning the content of experience. The arguments below are designed to deliver a fundamental principle at this depth, and while the principle will not contradict the axioms of IIT, by virtue of being metaphysically deeper its predictions are notably different, including two that are testable empirically.

## Section 2. Tackling the Hard Problem

Consciousness is typically defined in terms of what the world is like (Nagel, 1974); there is 'something it is like' to see redness, to feel pain, or to have thoughts and ideas. The 'it is like' aspect here captures the extreme privacy of subjective experience, so private and inaccessible that we can never be certain whether others experience it the same way, or even at all. There is nothing wrong with this as a description of the phenomenon, but as a formulation, it complicates matters unnecessarily, for the following reason. The description relies heavily on perceptual content, such as redness, pain, etc., and this has two confounding consequences, one to do with causality and other, with unconscious mind. On causality, On causaiity, there is no clearly defined boundary between objectively physical perceptual content and its qualitative aspects; the former has a physical cause, and the latter is part of the hard problem. The second confounding consequence of defining consciousness by content is that sensory and cognitive systems do their work unconsciously, and simply deliver content for conscious attention. It becomes conceivable to explain why red is red, or pain is painful, and still debate whether the explanation says more about unconscious perceptual systems than consciousness per se.

These complications can both be overcome with two assumptions. The first is that everything of which we are conscious is produced by physical systems and simply delivered to conscious attention. In other words, we will assume that every perceptible difference from nothing is governed by the principle of physical causality, including emotions, thoughts, inner dialogue, dreams, and the stuff we walk on. This assumption will free us to focus on the experiencer in whose presence physical reality exists. I am not the first to emphasise the central importance of presence for an explanation of consciousness (Franck, 2008; Seth et al., 2012, Seth 2014, 2015, Seager, 2019; Fasching, 2020).

The second assumption is that physical realism is true, together with its principle of causality. So, what do we mean by physical causality?

Causality is a first principle of empirical science: every observable phenomenon must have a cause. But despite being a first principle, some have argued that causality is not fundamental. One reason given is that causality always occurs locally whereas physical laws are universal. A second is that the equations that describe instances of causal action are usually time symmetrical whereas causality is asymmetrical; it has an arrow of time (Russell, 1913; Frisch,2012: Blanchard, 2016). A third possible reason is the requirement to define boundary conditions when describing actual causal events (Primas, 2002). Nevertheless, physical laws assume causation in its Humean sense: for any observed change from A to B such that A causes B, once A is given then B *must* happen. Causal relationships are not themselves observable, but are hypothetical inferences based on observations, and subject to falsification by further empirical data (Popper, 1953). Causation can never be proven true as a matter of principle. For any phenomenon with a causal explanation, it will remain forever possible that some aspect or example of the phenomenon will be observed for which the explanation does not hold.

For our purposes here, we will use a concept of physical causality sufficiently basic to be consistent with both relativity and quantum theory, with emphasis on the temporal sequencing of events. Spacetime is causally structured (Rahaman, 2021), and it is axiomatic in relativity theory that for any difference in spacetime between A and B, such that A causes B, A must be in the past light cone of B. The cause cannot be simultaneous with its effect. Once everything is known about A, it becomes possible to predict B with complete accuracy, and vice-versa. But what about quantum reality? In contrast to the complete predictability of relativity, quantum theory includes probabilities for predicting observable outcomes in the future. But even here it remains true that when A is known to have caused B, it must be in the past of B.

**Preliminary argument: Perception in simultaneous presence**[4]

The simultaneity being described here is purely phenomenological. It describes an experiencer being conscious when something is experienced. I use the words 'observer' and 'observed' in the same sense. We have already assumed that everything observed is subject to the laws of physics, but the nature of the conscious observer is yet to be explained, (the hard problem).

*The perception-continuity paradox:*

Spacetime is a four-dimensional continuum. The concept of continuity entails that *every non-zero point is really a tiny interval that always contains motion*. Each point is infinitely divisible into further intervals, which means that mass-energy, i.e., matter, is never static in time. Continuity may no longer apply at the Planck scale of $10^{-33}$ meters, and $10^{-44}$ seconds. But, by definition, the Planck scale specifies a limit, beyond which, nothing can be observed by any physical means. Movement is also ubiquitous in quantum physics, where objects as large as molecules can behave as waves during certain observations, such as the double-slit experiment. So, everything *observable* is moving continuously in time. With our two assumptions, everything observed by consciousness is physical and occupies some volume of spacetime in which there are more than two causally related temporal states. But for any two causally related parts of spacetime (A and B), there is no point where A and B can both be *present* in any temporal sense. No matter how small the interval between them, any one part of the causal continuum must always be in the past of a part it causes, which is a fundamental principle of causality and is an apparent paradox.

[4] Early versions of the perception argument have been published or presented previously, (Sanfey, 1999, 2003a, 2003b, 2005, 2020), but the subsequent, decisive argument is new.

- Every point in spacetime contains motion where some state A is continuously becoming some state B
- When motion is continuous, A and B can never be separated, every point contains both
- If A causes B, it cannot be present when B exists: a cause must be in the past of an effect it causes
- A exist as a reconstruction in the process of perception, call it $A_{\boldsymbol{R}}$, in relation to B

The principle of causality specifies that physical reality exists as A *then* B, but empirically it exists as A *and* B, where A must be a retention of A, (call it $A_R$), when B exists. No aspect of physical reality can exist in observable form, without some specific contribution by the process of perception. Physical reality has past and future aspects that are joined by the mind of the observer.

At first glance, there is nothing particularly new about this. Saint Augustine recognised in the fifth century that the conscious now contained past elements, as did William James who coined the term 'specious present' (James, 1890), and Edelman who coined the term 'remembered present' (Edelman, 1989). Husserl used the word 'retention' to describe the departing edge of now-ness, the just-now (Varela, 1999). However, the argument here serves to emphasise two important facts. Firstly, when matter is present, the past and future aspects of the spacetime it occupies, are joined together by the process of perception and not by any inner substance we may suppose matter to have. Secondly, there is no level of granularity of spacetime where this is not true.

Of course, the idea that the observer contributes something to the observed is not particularly new either. In predictive processing theory, perception helps shape the form in which matter is perceived (Friston, 2007, 2009, 2010, 2019; Friston & al, 2011, 2012; Seth, 2014; Kiefer, 2020). However, there is a further conclusion from the argument above, which to the best of my knowledge, is new. Any scientific conceptualisation of matter in science must also include and reflect the role played by a conscious observer's process of perception, for the following reason:

*The functional equivalence of now-ness:*

- For any perceived difference from nothing, the conceptual elements $A_{\boldsymbol{R}}$ and B are both present in the difference
- $A_{\boldsymbol{R}}$ and B are empirically inseparable
- Any *empirically testable* description of an observed difference from nothing must also reflect the process that links $A_{\boldsymbol{R}}$ and B, namely, the process of human perception
- Empirical reality always exists in a form that reflects the process of perception, whether it is presented in sensory or conceptual language

Theoretical science must overcome the same problem that evolution solved for conscious systems, namely, how to display past and future elements of time together. Our senses use the language of colour, texture, sound, etc., which enable physical reality to appear present beyond our boundaries. The problem of time has been equally difficult for science, especially when intervals become infinitesimally small yet still contain motion. It took more than two thousand years to solve Zeno's paradoxes of motion. Nowadays, we have tools such as calculus, field theory, configuration space, imaginary time, eigenvectors, and many more, that resolve various problems caused by continuity. But these tools are not physically real. We don't, for example, believe that calculus describes actual devices in the fabric of space-time that convert infinitesimal intervals into quantities to ensure that motion obeys the laws of physics. Perceptual systems produce sensory forms that span the interval of time that we call *now*, and the mathematical devices of science, along with the laws of physics, are abstract systems necessary to explain sensory perception. Since scientific models explain empirical observations, it follows that these abstract systems must be *functionally equivalent to the process of perception during empirical observation*. So, every perceptible difference from nothing exists in a form that is inseparable from the process of perception, both when experienced sensorily and when conceptualised in science.

- **Every difference from nothing contains an observer function that is functionally equivalent between sensory perception and conceptual description, and which appears intrinsic to the observed**

Matter is always and only knowable as some form of a model, whether sensory (subjective) or conceptual (objective). Both contain an observer function whose properties must be equivalent between the two perspectives. But because perceptual systems operate subconsciously, this principle of functional equivalence may say more about unconscious perception than consciousness per se. To tackle the hard problem proper, we need to focus on the experience of being present when perceptual content exists.

**The simultaneity argument Part 1: the limit of phenomenal certainty**

The next pair of arguments do not assume physical realism, nor anything else for that matter, and all inferences are deductive. If correct, the arguments will prove that the subjective experience of consciousness has causal power, irrespective of what we are conscious of.

It is worth repeating at this point, that the simultaneity in question is contained within any moment of experiencing. When we see something red, the redness is present simultaneously with the experiencer. In effect, the simultaneity between observer and observed is a more precise form of the hard problem, but because simultaneous causation is not possible in physical laws, the problem remains hard. A cause must be in the past of its effect, even for infinitesimal intervals of time. However, this formulation is perfectly solvable.

Before the arguments proper: a quick review of the rationale for both realism and idealism.

For a physical realist, it doesn't matter that reality is experienced in the sensory language of colours, textures etc. The important point is that perceived reality *behaves* independently of mind and follows the principle of physical causality without fail. Consequently, it makes sense to regard perceived reality as both physically real and causally closed. That typically leads to panpsychism and epiphenomenalism because unlike other intrinsic properties of matter such as electric charge or mass, consciousness is not associated with a fundamental force and should not be able to cause observable consequences within current physical laws.

An idealist or mentalist on the other hand, might agree that perceived reality demonstrates invariant features and follows the principles of physical causality with complete consistency. But that simply proves that reality behaves independently of *conscious awareness*, not unconscious mind. Since our experience of mind is both intimate and certain, it makes sense to an idealist to regard reality as the product of mind-stuff, even though we are not consciously aware of the processes involved.

Both interpretations of invariance are logically possible. Next, we take a step deeper and examine what can be said with complete certainty without any assumptions or inference.

Idealists and realists can agree that the following is empirically certain:

- Something is happening

They can also agree they are not consciously causing what is happening, for the simple reason that one cannot do something *consciously* without being *consciously aware* of it. Of course, we often become aware of actions that were unconscious a moment earlier, when driving a car for example, but these are not actions consciously initiated but rather unconscious ones that we subsequently become aware of. Both a solipsist and a physical realist will agree that the idea of doing something consciously without being consciously aware of it is philosophically a wooden iron, i.e., it is self-contradictory. Interestingly, this logic remains true without conscious content. Suppose the universe consisted of nothing, but that somehow, we were still conscious in some disembodied way. We would be aware of being aware, and further, we would know for certain that we were not *consciously* causing awareness to exist. If we were consciously causing our own awareness, we could stop causing it and instantly cease to exist.

We can now modify the list of statements that can be made with complete phenomenological certainty to:

1. There is something rather than nothing
2. It is happening now
3. We are not consciously causing it

Something interesting happens when we analyse how we know each of these to be true with complete certainty. The first two elements require no thought or any process that takes time to produce the necessary information, we simply know that we are experiencing. The third element on the other hand requires a moment of deductive thought, albeit a simple one. And, importantly, the premisses necessary for this deduction are the simultaneity between observer and observed. This innocuous sounding conclusion is the key to solving the hard problem, and it can be proven with deductive certainty.

**The simultaneity argument Part II: The causal power of conscious presence:**

The initial premisses for the next argument are firstly, to suppose that there is a super-smart artificial intelligence (AI), which has identical perception to humans, but does not experience conscious presence simultaneously with any content, and secondly, that it always tell the truth. The AI might realise that its perceptual systems could be inaccurate to the extent that everything it perceives might be fabricated entirely by its own systems. There is nothing it could learn or discover by trial-and-error that would prove it was not causing what it perceived. The reason is simple. An initial premiss was that the AI does not experience simultaneous presence, so its observing frame of reference must be its perceptual and cognitive systems. Once it becomes possible that perceived reality could be produced entirely by those systems, it is logically impossible to prove it is not causing the reality it perceives. A conscious observer on the other hand, can be certain they are not consciously causing reality because their observing frame of reference is simultaneously conscious. ***For a conscious observer, the simultaneous presence of awareness makes it logically possible that the observing frame of reference could be non-physical***. We can conceptualise and imagine the possibility of disembodied awareness, as we did above, but that is only possible because, unlike the AI, we experience being an observer simultaneously with the observed[5].

- When an observing frame of reference is not simultaneous with its content, it cannot be certain it is not causing the content

[5] The AI argument above has obvious similarities to Penrose's argument that consciousness must be non-computational and that algorithms merely follow decision rules, whereas consciousness can understand them (Penrose, 1989). In a conference session in 2019, he suggested that his argument might not hold true if there was '*an algorithm in our heads, which we don't know*' (Penrose, 2019, 18:30). This implies that for him at least, the problem is not with algorithms per se, but with the apparently never-ending need for a further algorithm. The simultaneity argument proves the same thing, but it can be interpreted as saying that observer-observed simultaneity enables us to recognise a circular argument when we see one, unlike a computer.

We now have sufficient means to resolve the hard problem. The hypothesis that mind-stuff could exist as a distinct class of substance, or even as the only substance, becomes logically possible when both observer-observed simultaneity and the fact that the observer is not simultaneously causing reality, are both known to be true with complete certainty. In other words, simultaneity makes some explanations logically possible that would otherwise be impossible. Ideas can influence our behaviour, not least by causing pause for thought, so this *establishes as fact that conscious presence is sufficient to cause behavioural change in the physical world, irrespective of anything it may be conscious of*. The point may seem somewhat trivial, but it breaks the stranglehold of the hard problem and establishes a connection between consciousness and physics at the most fundamental level. With proven causal power, it follows that consciousness can be integrated with the same science as physical causality, and further, that it can be produced by physical systems.

Key points in simultaneity argument:

- **When an observing frame of reference is simultaneous with the observed, and not related to it by simultaneous causation, it becomes logically possible that the observer is of different substance to the observed**
- Conceptual alternatives to physical realism are not logically possible without observer-observed simultaneity
- Concepts can influence behaviour
- Consciousness has causal power

The next section collects the principles established above and combines them into a single bridging principle with an explanation of how conscious causality operates without breaking the laws of physics. It is followed by a section describing what that principle predicts in terms of a physical theory of phenomenal consciousness.

## Section 3. Metaphysical principle & testable hypotheses

In summary, the two principles established by the arguments above are as follows:

1. Every difference from nothing contains an observer function that is functionally equivalent between sensory perception and conceptual description, and which appears intrinsic to the observed
2. Observer-observed simultaneity creates logical possibilities that could not exist without it, even when the simultaneity is between experiencing and awareness of the fact that experiencing is happening

In the following paragraphs I will consolidate these ideas into a formal bridging principle within a conceptual framework that I call *Abstract Realism*.

### Abstract Realism (AR)

Every difference from nothing, whether subjectively experienced or objectively described, contains an observer function whose purpose is to project the form and meaning necessary to explain what is happening, and which is functionally equivalent in both subjective experience and the theoretical models of science. The observer function is best conceptualised as a virtual dimension that should be considered physically real despite being imperceptible when experienced as conscious presence and abstract in the concepts we infer from experience. In both cases the dimension is epistemic in the sense that it evolved to provide the structure that enables living systems to challenge and explain what triggers perceptual content in the first place.

Science has acted as if it can describe matter without an observer. It cannot. Every point in spacetime has past and future aspects connected by an *intrinsic* observer function. The 'something it is like' of experiencing redness, hardness, or heaviness has a functional equivalent in electromagnetic radiation, electrostatic forces, and the spacetime curvature of gravitation where abstract devices, rules, laws, and boundary conditions describe and quantify the behaviour of observed phenomena over time. There is always an invisible, epistemic frame of reference projecting the form and meaning required for the critical evaluation and quantification of uncertainty, and this is true for any interval of time, however small. The observer function in physics can be structured in any way we like so long as it works (Sanfey, 2005), because it is functionally equivalent to something imperceptible, namely, a conscious observer. The observer function is necessary for matter to have difference from nothing. But it also has causal consequences, which are actioned by the same mechanism in both consciousness and science.

Matter always exists as some form of model, whether sensory or conceptual. The sensory models of redness, hardness, or heaviness become conceptual as electromagnetic waves, electrostatic forces, and the spacetime curvature of gravitation. Modelling is fundamental throughout biology, from the receptor molecules in cell membranes to the theoretical models in physics. As scientific method developed over time, so sensory perception evolved through millennia by natural selection of the fittest. Both forms of modelling try to predict how the external environment will behave. But in both cases, matter is indifferent to our models. It behaves the same whatever we think about it. There is never any simultaneous interaction between observer and observed. But there is interaction nonetheless, and it occurs by the same mechanism in both phenomenal consciousness and theoretical physics. Let me explain.

Science operates by the predictive modelling of theories such as quantum mechanics and relativity. Sometimes the predictions do not match what is observed, and the theory or model in question is considered incomplete. This happened to Newtonian mechanics when Maxwell's equations of electromagnetism predicted that the speed of light in a vacuum was constant for all observers, an anomaly the Newtonian model could not explain. Eventually, after several decades, Einstein's theories of relativity were developed, and epistemic order was restored to classical physics. But theories themselves do not influence how matter behaves. The meaning of electricity may change for humanity, or the meaning of time, gravitation or matter itself, but none of that affects the way particles or planets behave. What changes is the behaviour of people interpreting theoretical models and using their predictive power to make decisions in the real world. This exactly parallels how consciousness operates the world it inhabits: the world of sensory models and conceptual thought. Just as matter is indifferent to any change in scientific laws or theory, it is also indifferent to the conscious meaning of what we perceive. We have no power to influence reality simultaneously by an act of will, but the way we interpret perceptual content will change our future actions in the world being modelled, and those actions proceed by physical movement in accordance with physical laws. This is how abstract realism explains the causal power of experiencing presence in time without breaking existing laws.

*The Bridging Principle:*

There is an important ontological twist to the intrinsic observer-observed relationship that must be completed before stating the bridging principle. When sensory models exist, a conscious observer experiences their own presence as real but cannot be certain whether reality is mind-stuff or independently physical. In the science of physical realism, the opposite ontology applies. Here matter is considered real with its principle of closed physical causality, and the observer is ontologically abstract in the form of laws, devices, mathematics, etc. The observer-observed relationship is functionally equivalent in both cases, subjective and objective, but with ontological reversibility in terms of which is real, and which is abstract.

Observer-observed duality is best described by *complementarity,* a concept introduced by William James (1890) and developed by Bohr to accommodate apparent wave-particle duality of matter at the quantum level (Bohr, 1927). Inspired by Heisenberg's more formal uncertainty principle, Bohr's complementarity describes circumstances where two mutually contradictory explanations are both necessary for the complete classical description of a microscopic system. Wave and particle explanations appear mutually contradictory, but both are necessary for a complete description of observed behaviour in the macroscopic world of perceived reality. In terms of the hard problem, observer and observed appear mutually exclusive, but in AR each is present, both at the phenomenal level of subjective experience and the conceptual level of thought and science, with the ontological twist between subjective and objective perspectives described above. If there were no observer function in either perspective, no intervals of spacetime would exist, there would be no movement, no order or meaning and every aspect of reality would be indistinguishable from nothing.

Observer-observed duality is metaphysically fundamental. It appears immediately we reflect on experience, and it defines the boundary that limits empirical certainty. As soon as we realise that there is something not caused by conscious presence, uncertainty is created, and a conceptual level of reality comes into existence, with its own intrinsic observer-observed duality and opposite ontology. The relationship between subjective and objective levels of reality is a complementarity pairing of two perspectives, each with its own observer-observed complementarity. In both subjective and objective perspectives, the observer function is functionally equivalent.

The bridging principle of consciousness arising from the arguments is now complete, and can be stated as follows:

> **In any difference from nothing, whether subjectively experienced or objectively described, there is an observer-observed relationship such that the observer is functionally equivalent but ontologically opposite in the subjective and objective perspectives, and whose function is intrinsic to the observed but never its cause, and which can always create an additional degree of uncertainty concerning the nature of the observed.**

Just as the phenomenological axioms of IIT have corresponding postulates, this bridging principle postulates how consciousness can be produced by physical systems provided we use the term 'physical' with equal meaning for both idealism and realism. 'Physical' here describes observed behaviour over some interval of time, such that past and future aspects of observed behaviour acquire form and meaning from the observer function which appears intrinsic within the observed. Physical reality contains an epistemic dimension which is the observing frame of reference that defines its observable features. But the observer itself is invisible to the observed. In the case of consciousness, the observer has no 'physically' observable features, and in physics, it is entirely abstract. Let us now examine what this predicts in terms of brain function.

*AR and the brain:*

There is good evidence that connectivity between *feedforward* and *feedback* modifier signals is necessary for consciousness and that feedforward signals from sensory and cognitive stimuli, propagate through various content relevant areas of the brain during consciousness (Koch et al., 2016). In both IIT and GW theories the degree to which a system is conscious is the extent to which any point at the interface of interacting systems can create rapid and wide access to distributed system memory, or cause action by the whole system. These ideas are all compatible with AR, but they fall short for explaining the phenomenal simultaneity between observer and observed required by AR's bridging principle.

The problem of simultaneity is closely related to the long-standing *combination problem* of consciousness often attributed to James (1890). This is the problem of explaining how the brain combines a rich tapestry of multimodal perceptions and thoughts into a unified phenomenal field. The bridging principle of AR provides the basis of a solution. AR predicts that consciousness is phenomenally instantiated when two EM fields interact causally with the substrate generating the other. The postulate is that information in the EM fields created by synchronised neuronal firing tries to mirror itself in system memory and that this is achieved when its EM field resonates with similar EM fields multiply created in the microtubules (MTs) within the firing neurons. This hypothesis may seem far-fetched, but there is increasing evidence to suggest it is feasible.

Microtubules (MTs) are tubular structures that exist in all eukaryotic cells, namely those cells that contain internal, membrane-bound organelles. MTs are important for transport, cell division and construction, but also for memory in the plasticity of dendritic connections (Dent, 2017). It has been suggested that the symmetrical arrangement of the tubulin building blocks of MTs may also be important (Sahu et al., 2013b; Janke et al., 2014, 2020; Hameroff, 2022). MTs are likely to play some essential role in consciousness because they are highly sensitive to general anaesthetics that switch consciousness off during surgical procedures (Craddock et al., 2015). AR's postulate predicts that their electromagnetic (EM) properties are the critical component for consciousness.

There is evidence that MTs can both store EM information and then transmit it electronically. Kalra and colleagues have demonstrated long-lived, collective behaviour within MTs enabling them to 'harvest light' effectively, (Kalra et al., 2023). MTs have been shown to both produce their own EM fields (Pokorny et al., 2021) and to interact with others, (Sahu et al., 2014; Gholami et al., 2019; Saxena et al., 2020). They can generate electrical impulses that are similar in nature to the action potentials of neural axonal firing (Del Rocío Cantero et al., 2018). Bandyopadhyay's team published evidence suggesting that MTs can provoke and regulate axonal firing (Singh et al., 2021a), and also evidence of EM energy exchange between MTs and neurons at the KHz ($10^{-3}$s), MHz ($10^{-6}$), and even GHz ($10^{-9}$) scales, (Singh et al., 2021b). Tuszynski recently described some yet to be published work by Dogariu, which shows that MTs can release light for up to *fifteen minutes* after interacting with light and other electromagnetic radiation, (Tuszynski, 2021, 41:05). The current world record storage time for coded light information in non-biological systems stands at one hour (Ma & al, 2021). Despite widespread quantum scepticism, (Tegmark, 2000; Koch, 2006), molecular quantum effects are undoubtedly important throughout biology. They are essential for photosynthesis, for enzyme catalysis, for the magnetoreception used by migrating birds for navigation, (Lambert et al., 2013), and for the complex folding of large molecules.

AR's biological postulate is that consciousness is instantiated in the bi-directional interaction between synchronised neural firing and system memory reported by microtubules (MTs). The observing frame of reference may be reversible between the two EM fields, but normally would be from system memory, where previous firing patterns are accessed and re-transmitted quantum mechanically by MTs[6]. The EM field caused by synchronous neural firing seeks resonant patterns with system memory as it produces the MTs' EM fields. The speciousness of the conscious now is explained by the fact that the two mirrored information systems exist in different media, at vastly different scales, at different phases in time with different durations of existence but communicate with each other's substrate using similar EM fields. All these factors contribute to the thickening of time we experience as the conscious now, in a similar manner to Edelman's theory of neuronal group selection (TNGS), whereby synchronously and recursively firing neuronal groups across widely distributed brain areas are selectively favoured (Edelman 2003), although Edelman's TNGS does not have an EM component.

[6] Bergson also described awareness as observing from the perspective of memory but he claimed that memory was a spiritual substance and not directly involved in perception, except to modify percepts already present (Bergson, 1896). Bergson never tried to explain consciousness except to say that it must be described in temporal terms. The issue that eroded his credibility the most, was simultaneity. In a famous debate with Einstein, he defended the argument that a moment of simultaneity between observed events was real and the same for all beings (Bergson, 1922).

AR's biological postulate requires shorter coherence times than those quantum mind theories which have more extensive quantum ontologies. The quantum states of individual MTs interacting with synchronously firing neurons can be refreshed many times during the 100 or so milliseconds of the conscious now ($10^{-1}$ seconds). In that way, this model addresses Tegmark's objections (Tegmark, 2000), and in any case, there is increasing evidence that the unique structure of MTs supports much longer coherence times than those with occur in the other quantum effects throughout biology (Sahu et al., 2013a).

The theory predicts that as the study of the brain's EM fields develops, evidence of strong correlation will be found in both intensity and pattern between the EM transmissions of microtubules, most likely at the KHz scale ($10^{-3}$s), with the EM fields in the produced by neural synchrony at the Hz scale. The theory also predicts that this correlation will itself correlate with the conscious experience of test subjects.

A further prediction of AR is worth mentioning here although it originates in the simultaneity argument rather than the biological predictions of the bridging principle. Recent versions of IIT state categorically that there can be no Turing type test for consciousness (Tononi and Koch, 2015), although earlier versions suggested that a such a test was possible. (Koch & Tononi, 2008). In AR on the other hand, the feasibility of such a test is strongly predicted based on the premiss of the simultaneity argument, in which an Artificial Intelligence (AI) that does not experience simultaneous presence, but which has knowledge of its own function, and no prior knowledge of how a conscious system would answer the test, will be unable to prove it is not causing its own perceptual content.

**Discussion**

If the arguments here are truly deductive, the hard problem is resolved by proof that consciousness can create conceptual content in any moment of time. Even if the universe were nothing except for our own consciousness, we could still create competing hypotheses to explain why consciousness exists; it could either be created by our own unconscious mind or by completely independent of ourselves. Either way, we would know for certain that we were not *consciously* causing consciousness. Being conscious is sufficient to create uncertainty and conceptual choice, irrespective of any content. At the very least, uncertainty can cause pause for thought, and concepts shape actions. This establishes the principle that conscious presence has causal power.

Of course, the universe is not nothing, and we do experience a rich and varied content. The bridging principle makes it possible for consciousness to be explained in the same science that describes that content, without breaking existing laws of physics. The principle can be stated as: **in any difference from nothing, whether subjectively experienced or objectively described, there is an observer-observed relationship such that the observer is functionally equivalent but ontologically opposite between the subjective and objective perspectives, and whose function is intrinsic to the observed but never its cause, and which can always create an additional degree of uncertainty regarding the nature of the observed.**

The hard problem could never be solved by any psycho-physical identity because a principle governing the mind-matter relationship must include the choice to infer physical realism, a choice that has causal consequences. Instead of a psycho-physical identity AR describes a deeper subjective-objective complementarity that applies to any difference from nothing, including the realisation that our own consciousness is not nothing in terms of existence. **For any difference from nothing there is an observer; the observer always has a choice, and that choice always has causal consequences.** Consciousness generates two perspectives, subjective and objective, experiential and conceptual, and each perspective contains its own observer-observed relationship. In the subjective perspective the observed is modelled in the language of sensory perception, and once we think about it conceptually, it is modelled in the science of physical realism. In both subjective and objective perspectives, the observer is functionally equivalent but with reversed ontology. When sensory models exist, the observer is our conscious presence, and when the models are conceptual, it exists in a variety of abstract equivalents. In both cases, the observer function is to project the form and meaning necessary to display and evaluate perturbations in our predictive systems through time.

Mind and matter belong in the same science, provided we use the term 'physical' with equal meaning for both idealists and realists. We can never know what matter really is, as Kant, Schopenhauer, and others have pointed out. Matter is always a model, and in both subjective experience and theoretical science, the causal interaction between humans and their models is by the same mechanism; when a model is replaced, or its meaning changes, the behaviour of matter remains the same, but human behaviour changes.

There are some notable differences between the predictions of AR and f IIT, some of which are empirically testable. Whereas IIT explicitly rules out the possibility of a Turing type test for consciousness (Tononi & Koch, 2015), the simultaneity argument in AR is the basis for such a test. Only a system that is *simultaneously* aware of its own observing presence can pass AR's test, and any system that does pass must be considered conscious. Another difference is that in AR it is impossible to measure $\Phi$ as a matter of principle because intrinsic simultaneity is what creates the additional degrees of causal freedom possessed by consciousness, and extrinsic observation cannot be simultaneous but must entail some temporally sequential process. Mørch was correct to point out that IIT's $\Phi^{max}$ cannot be both subjectively intrinsic and measurable extrinsically (Mørch 2019). This is the same fundamental problem that forced IIT and FIIH to reach first for panpsychism before pivoting towards a 'causal emergence' ontology.

The arguments here prove two things that are necessary for a causal emergence ontology of consciousness. Firstly, consciousness has causal properties, and secondly, its intrinsic causal power is unobservable in principle because it results from simultaneity within the phenomenon. With proven causal power, it becomes possible for consciousness to be produced by physical systems, and by being unobservable in principle, it is no longer necessary for a theory to explain the causal freedom of concsiousness at the level of fundamental fields or forces, because that freedom is proven (if correct) to both exist and be unobservable in principle. The field of causal emergence can be liberated from metaphysical constraints and focus on technical aspects, (Mediano et al., 2022a, 2022b; Albantakis et al., 2022, 2023).

Various hypotheses are possible for how systems might produce consciousness based on AR's bridging principle. The theory favoured here is that consciousness is produced when two substrates interact bi-directionally via their respective EM fields, namely the synchronously firing neurons, and the microtubules within those neurons. The hypothesis is testable by its prediction of strong correlation between the bi-directional, mirroring EM fields, though not necessarily at the same frequency. The technology for testing this prediction is rapidly becoming available (Singh et al., 2021a, 2021b, 2021c).

The importance of modelling in AR does not imply that information is ontologically fundamental in the manner described by IIT or FIIH. In AR, reality looks informational because our systems are trying to sense of it and keep us alive. AR's concept of information has more in common with Wheeler's 'it from bit' theory of fundamental reality than with IIT. For Wheeler, everything physical derives its existence by answering questions posed by a measuring instrument; '*all things physical are information-theoretic in origin, and this is a participatory universe*' (Wheeler, 1990).

We experience the observer function as being a negative space of potentially infinite meaning, where the term 'negative space' is not derogatory in any sense but captures the fact that consciousness has no spacetime dimensions and is physically invisible: abstract realism. In an evolutionary context, consciousness is a highly sensitive system for evaluating subtle nuance and barely perceptible pieces of information that bubble in the periphery of awareness, very much in line with Libet's experiments. Libet pointed out that while actions can be initiated prior to conscious awareness, the action itself always occurs some 200ms after conscious awareness (Libet et al., 1983), meaning that consciousness retains the power of veto over unconsciously initiated actions. For that reason, he claimed that his experiments were compatible with the concept of free will. In AR, consciousness depends heavily on the unconscious, but can create uncertainty in a flash. A hint of conscious scepticism creates time for millions of bits of information to interact unconsciously in the brain enabling new ideas to compete and prepare for action.

I cannot speculate on the mathematics required to formalise the observer-observed interaction in AR's bridging principle. Others may be able to do so and make further testable predictions as a result. The approach here concerns principles. However, in AR there is always uncertainty at the observer-observed interface which the field of FEP-AI seems well suited for since it describes boundaries as being necessarily fuzzy with weak coupling between interacting adaptive systems (Ramstead et al., 2023). But other approaches are also possible, especially Vitiello's dissipative quantum field theory, whose description of open systems also requires a mirroring system (Vitiello, 1995, 2005; Freeman & Vitiello). It may also be the case that the weak quantum measurement approaches described by Aharanov and colleagues (2021, 2022) may prove important. And, as noted earlier, the IIT team have responded to Barrett and Mediano's critique and are adapting their mathematical models to incorporate a possible quantum ontology (Albantakis, 2023b), as are others (Zanardi et al., 2018). IIT is not falsified by AR, but is simply shown to be not fundamental.

In conclusion: the simultaneity between observer and observed is both the underlying cause of the hard problem and its solution. The simultaneity exists for certain in consciousness, but causation cannot be simultaneous. The key to solving the riddle is that matter always exists as a model, whether sensory or conceptual, and models are epistemic. They come with an implicit observer. The intrinsic observer function determines the meaning of models in both cases, and when the meaning changes, the future actions of humans using the models may change too. When models are written in the language of sensory perception, the simultaneous conscious presence can exert causal power by seeing different meaning. By possessing causal power, consciousness can be considered physical despite being physically imperceptible.

Lastly, simultaneity is responsible for the extreme privacy of consciousness. We cannot even see our own presence without looking *from* presence. Consciousness cannot be detected by anything that requires movement over time. Conscious presence is certain, but only when experienced simultaneously with reality. Simultaneity ensures that presence can generate infinite scepticism in a manner impossible to observe or predict by any means. If correct, this proves that we do have free will in a physical universe, and that consciousness has evolved for a good reason: it makes us contrary.

**References:**

Aharonov, Y., Colombo, F., Sabadini, I., Shushi, T., Struppa, D. C., & Tollaksen, J. (2021). A new method to generate superoscillating functions and supershifts. Proceedings of the Royal Society a-Mathematical Physical and Engineering Sciences, 477(2249), Article 20210020. https://doi.org/10.1098/rspa.2021.0020

Aharonov, Y., Colombo, F., Jordan, A. N., Sabadini, I., Shushi, T., Struppa, D. C., & Tollaksen, J. (2022). On superoscillations and supershifts in several variables. Quantum Studies-Mathematics and Foundations, 9(4), 417-433. https://doi.org/10.1007/s40509-022-00277-x

Albantakis, L., Barbosa, L., Findlay, G., Grasso, M., et al (2022). Integrated Information Theory (IIT). 4.0: Formulating the properties of phenomenal existence in physical terms. *arXiv:*2212.14787.

Albantakis, L.; Prentner, R.; Durham, I. (2023). Measuring the integrated information of a quantum mechanism. *Preprints*, *1*, 0. https://doi.org

Anastassiou, C. A., Perin, R., Markram, H., and Koch, C. (2011). Ephaptic coupling of cortical neurons. *Nat. Neurosci.* 14, 217–223. doi: 10.1038/nn.2727

Armstrong, D., (1968) *A Materialist Theory of Mind.* London: Routledge and Kegan Paul,

Aru, J, Bachmann, T. Singer, W. Melloni, L. (2012), Distilling the Neural Correlates of Consciousness, *Neuroscience and Biobehavioral Reviews*, 36(2): 737–746. doi: 10.1016/j.neubiorev.2011.12.003

Atasoy, S., Donnelly, I., Pearson, J., (2016). Human brain networks function in connectome-specific harmonic waves. Nat Commun. 7:10340. doi: 10.1038/ncomms10340. PMID: 26792267; PMCID: PMC4735826

Atmanspacher, H., (2004/2020). Quantum Approaches to Consciousness, *The Stanford Encyclopedia of Philosophy*. Edward N. Zalta (ed.), URL = https://plato.stanford.edu/archives/sum2020/entries/qt-consciousness/

Atmanspacher, H., (2014), 20th century variants of dual-aspect thinking (with commentaries and replies), *Mind and Matter*, 12: 245–288.

Baars, B., (1988). *A cognitive theory of consciousness*. New York, NY: Cambridge University Press.

Baars, B., (1997). *In the Theater of Consciousness* (New York, NY: Oxford University Press)

Baars, B., (2002). The conscious access hypothesis: Origins and recent evidence. *Trends in Cognitive Sciences,* 6 (1), 47-52.

Baars, B., Franklin, S., and Ramsøy, T. (2013). Global workspace dynamics: cortical "binding and propagation" enables conscious contents. *Front. Psychol.* 4:200. doi: 10.3389/fpsyg.2013.00200

Baars, B., and Geld, N. (2019). *On Consciousness: Science and Subjectivity—Updated Works on Global Workspace Theory*. New York, NY: The Nautilus Press Publishing Group.

Baars, B., Geld, N. Kozma, R. (2021). Global Workspace Theory (GWT) and Prefrontal Cortex: Recent Developments*. Front. Psychol. 12:749868. doi: 10.3389/fpsyg.2021.749868*

Barrett, A., Seth, A., (2011) Practical Measures of Integrated Information for Time-Series Data. *PLoS Comput Biol* 7(1): e1001052. https://doi.org/10.1371/journal.pcbi.1001052

Barrett, A., (2014). An integration of integrated information theory with fundamental physics. *Front Psychol.*; 5:63. doi: 10.3389/fpsyg.2014.00063. PMID: 24550877; PMCID: PMC3912322.

Barrett, A., (2016). A comment on Tononi & Koch (2015) 'Consciousness: here, there and everywhere?' *Phil. Trans. R. Soc. B.* 371: 20140198. http://dx.doi.org/10.1098/rstb.2014.0198

Barrett, A., Mediano, P., (2019). The Phi measure of integrated information is not well-defined for general physical systems*. J Conscious Stud. 21: 133. doi:10.1021/acs.jpcb.6b05183.s001i*

Bayne, T., (2018). On the axiomatic foundations of the integrated information theory of consciousness. *Neuroscience of Consciousness*: niy007. doi:10.1093/nc/niy007

Bergson, H. (1896). Matter and Memory. Translated by Nancy Margaret Paul and W. Scott Palmer. London: George Allen and Unwin (1911)

Bergson, H. (1922). *Duration and Simultaneity*., trans. Leon Jacobsen, with an Introduction by Herbert Dingle, Library of Liberal Arts (Indianapolis: Bobbs-Merrill, 1965).

Blanchard, T., (2016). Physics and causation. *Philosophy Compass* 11, 256-266.

Bohm, D., (1980), *Wholeness and the Implicate Order*, Routledge. London. E-book: http://www.gci.org.uk/Documents/DavidBohm-WholenessAndTheImplicateOrder.pdf

Bohm, D., (1990). A new theory of the relationship of mind and matter. *Philosophical Psychology, 3*(2), 271–286.

Bohm, D., Hiley, B., (1993): *The Undivided Universe: An Ontological Interpretation of Quantum Theory*, Routledge, London.

Bohr, N., (1927). The Quantum Postulate and the Recent Developments of Atomic Theory. Atti del Congresso dei Fisici, Como, 11-20 Settembre, Zanicelli, Bologna

Broad, C., (1925). *The Mind and its Place in Nature.* London: Routledge.
Carruthers, P., (2005). *Consciousness: Essays From a Higher-Order Perspective*. Oxford, GB: Oxford University Press UK
Chalmers, D., (1995), Facing up to the problem of consciousness. *Journal of Consciousness Studies,* 2 (3) doi: 10.1093/acprof:oso/9780195311105.003.0001
Chalmers, D., (1996), *The conscious mind.* Oxford. Oxford University Press.
Chalmers, D., (2006). *Strong and weak emergence*. In The re-emergence of emergence. P. Clayton and P. Davies, eds. Oxford University Press). http://consc.net/papers/emergence.pdf
Chiang, C., Shivacharan, R., Wei, X., Gonzalez-Reyes, L., Durand, D., (2019). Slow periodic activity in the longitudinal hippocampal slice can self-propagate non-synaptically by a mechanism consistent with ephaptic coupling; *The Journal of physiology* 597 (1), 249-269. DOI: 10.1113/JP276904
Clark, A., (2013). Whatever next? Predictive brains, situated agents, and the future of cognitive science. *Behavioral and Brain Sciences,* 36(3), 181–204.
Craddock, T., Hameroff. S., Ayoub, T., Klobukowski, M., Tuszynski, J., (2015) Anesthetics Act in Quantum Channels in Brain Microtubules to Prevent Consciousness*. Current Topics in Medicinal Chemistry, 15, 523-533*
Davidson, D., (1970). *Essays on Actions and Events*: Philosophical Essays Volume 1. Oxford, GB: Clarendon Press.
De Graaf, T. de, Hsieh, P. Sack, A. (2012), "The 'Correlates' in Neural Correlates of Consciousness", *Neuroscience & Biobehavioral Reviews*, 36(1): 191–197. doi:10.1016/j.neubiorev.2011.05.012
Dehaene, S., Changeux, J., Nadal, J., (1997). Neural networks that learn temporal sequences by selection. *Proc Natl Acad Sci* U S A. May;84(9):2727-31.
Dehaene, S., and Changeux, J., (2011). Experimental and theoretical approaches to conscious processing*. Neuron* 70, 200–227. doi: 10.1016/j.neuron.2011.03.018
Dehaene, S., Sergent, C., Changeux, J. P. (2003) A neuronal network model linking subjective reports and objective physiological data during conscious perception. Proc Natl Acad Sci U S A 100, 8520-5. https://doi.org/10.1073/pnas.1332574100
Del Rocío Cantero, M.; Etchegoyen, C.V.; Perez, P.L.; Scarinci, N.; Cantiello, H.F. (2018). Bundles of Brain Microtubules Generate Electrical Oscillations*. Sci. Rep., 8, 11899.*
Dennett, D., (1996). Facing backwards on the problem of consciousness*. Journal of Consciousness Studies.* **3** (1): 4–6.
Dennett, D., (2016) Illusionism as the obvious default theory of consciousness. *J Conscious Stud*; 23, No. 11–12, 2016, pp. 65–72
Dennett, D., (2018). Facing up to the hard question of consciousness. *Philos Trans R Soc Lond B Biol Sci*;373(1755):20170342. doi: 10.1098/rstb.2017.0342. PMID: 30061456; PMCID: PMC6074080.
Dent, E., (2017). Of microtubules and memory: implications for microtubule dynamics in dendrites and spines. *Mol Biol Cell*. Jan 1;28(1):1-8. doi: 10.1091/mbc.E15-11-0769. PMID: 28035040; PMCID: PMC5221613.
Edelman, G., (1989) *The Remembered Present: A Biological Theory of Consciousness.* Basic Books, New York.
Edelman. G., (2003). Naturalizing consciousness: a theoretical framework. *Proc Natl Acad. Sci. U S A.* 100(9):5520-4. doi: 10.1073/pnas.0931349100.
Fasching, W. (2020). Prakāśa. A few reflections on the Advaitic understanding of consciousness as presence and its relevance for philosophy of mind. *Phenom Cogn Sci* https://doi.org/10.1007/s11097-020-09690-2
Feinberg, T., Mallatt, J., (2020). Phenomenal Consciousness and Emergence: Eliminating the Explanatory Gap.*Frontiers in Psychology.* 11: DOI=*10.3389/fpsyg.2020.01041*

Fingelkurts, A. Fingelkurts, A. Neves, C. F. H. (2009). Phenomenological architecture of a mind and operational architectonics of the brain: the unified metastable continuum. *New Math. Nat. Comput. 5*, 221–244. doi: 10.1142/S1793005709001258

Fingelkurts, A., Fingelkurts, A., Neves, C. (2013). Consciousness as a phenomenon in the operational architectonics of brain organization: criticality and self-organization considerations. *Chaos Solitons Fractals* 55, 13–31. doi:10.1016/j.chaos.2013.02.007

Franck, G. (2008). Presence and reality: An option to specify panpsychism? *Mind and Matter*, 6(1):123–140.

Frankish, K., (2016). Illusionism as a theory of consciousness. *J Conscious Stud*; 23:11–39

Freeman W.; Vitiello G. (2006). Nonlinear brain dynamics as macroscopic manifestation of underlying many-body dynamics. *Physics of Life Reviews*. **3** (2): 93-118. doi:10.1016/j.plrev.2006.02.001. S2CID 11011930.

Frisch, M. (2012) No place for causes? Causal skepticism in physics. *Euro Jnl Phil Sci* 2: 331-336.

Friston, K., Stephan, K., (2007). Free-energy and the brain. *Synthese* 159, 417–458.

Friston, K., (2009). The free-energy principle: A rough guide to the brain? *Trends in Cognitive Science,* 13(7), 293–301.

Friston, K., (2010). The free-energy principle: a unified brain theory? *Nat. Rev. Neurosci.* 11, 127–138. doi: 10.1038/nrn2787

Friston, K., Mattout, J., Kilner, J. (2011). Action understanding and active inference. *Biological Cybernetics,* 104(1–2), 137–160.

Friston, K.J., Thornton, C., and Clark. A. (2012). Free-Energy Minimization and the Dark-Room Problem. *Frontiers in Psychology 3: 130.*

Friston, K., (2019). A free energy principle for a particular physics. *ArXiv190610184* Q-Bio. Available online at: http://arxiv.org/abs/1906.10184

Frohlich, F., and McCormick, D. A. (2010). Endogenous electric fields may guide neocortical network activity. *Neuron.* 67, 129–143. doi: 10.1016/j.neuron.2010.06.005

Gennaro, R. (1993). Brute Experience and the Higher-Order Thought Theory of Consciousness. *Philosophical Papers* 22: 51-69.

Gholami, D., Riazi, G., Fathi, R. *et al.* (2019). Comparison of polymerization and structural behavior of microtubules in rat brain and sperm affected by the extremely low-frequency electromagnetic field. *BMC Mol and Cell Biol* **20,** 41 (2019). DOI: 10.1186/s12860-019-0224-1

Hameroff, S., Penrose R., (1996), Conscious events as orchestrated space-time selections. *Journal of Consciousness Studies;* 3, pp 36-53 DOI: 10.14704/nq.2003.1.1.3

Hameroff, S., (2006). *Consciousness, neurobiology and quantum mechanics*, in: Jack A. Tuszynski (5 September 2006). The Emerging Physics of Consciousness. Springer Science & Business Media. pp. *192–251.*

Hameroff, S., (2022). Consciousness, Cognition and the Neuronal Cytoskeleton - A New Paradigm Needed in Neuroscience. *Front Mol Neurosci*. Jun 16;15:869935. doi: 10.3389/fnmol.2022.869935. PMID: 35782391; PMCID: PMC9245524.

Hales C. G. (2014). The origins of the brain's endogenous electromagnetic field and its relationship to provision of consciousness. *Journal of integrative neuroscience*, *13*(2), 313–361. https://doi.org/10.1142/S0219635214400056

Hales, C. Ericson, M. (2022) Electromagnetism's Bridge Across the Explanatory Gap: How a Neuroscience/Physics Collaboration Delivers Explanation Into All Theories of Consciousness. *Front. Hum. Neurosci.* 16:836046. doi: 10.3389/fnhum.2022.836046

Harris, S., Field, E., Imamoglu, A. (1990). Nonlinear optical processes using electromagnetically induced transparency. *Physical review letters*, *64*(10), 1107–1110. https://doi.org/10.1103/PhysRevLett.64.1107

Helmholtz, H., (1866/1962). Concerning the perceptions in general, in *Treatise on Physiological Optics*, 3rd Edn, Vol. III, ed. J. Southall, trans. (New York: Dover).

Hiley, B.J., (2001), Non-commutative geometry, the Bohm interpretation and the mind-matter relationship, in *Computing Anticipatory Systems—CASYS 2000*, D. Dubois (ed.), Berlin: Springer, pp. 77–88.

Hoel, P., Albantakis, L., Tononi, G., (2013). Quantifying causal emergence shows that macro can beat micro. *Proceedings of the National Academy of Sciences* 110, 19790–19795.

Hoel, E., Albantakis, L., Marshall, W., Tononi, G., (2016). Can the macro beat the micro? Integrated information across spatiotemporal scales. *Neuroscience of Consciousness 10 (49) 19790-19795 https://doi.org/10.1073/pnas.1314922110*

Hoel, E., (2021). A primer on causal emergence: Reply to Scott Aaronson. *The Intrinsic Perspective.* https://www.theintrinsicperspective.com/p/a-primer-on-causal-emergence (accessed, 31.7.23)

Hunt, T. (2011). Kicking the psychophysical laws into gear a new approach to the combination problem. *J. Conscious. Stud.* 18, 11–12.

Hunt, T., and Schooler, J. W. (2019). The easy part of the hard problem: a resonance theory of consciousness. *Front. Hum. Neurosci.* 13, 378. doi: 10.3389/fnhum.2019.00378

Jackson, F., (1982). Epiphenomenal Qualia. *Philosophical Quarterly* 32 (April): 127-136.

Jackson, F., (1986). What Mary Didn't Know. The Journal of Philosophy 83 (5): 291-295.

James, W. (1890) *The Principles of Psychology,* Volume One, New York: Holt.

Janke, C. (2014). The tubulin code: molecular components, readout mechanisms, and functions. *J. Cell Biol. 206, 461–472*. doi: 10.1083/jcb.201406055

Janke, C., and Magiera, M. M. (2020). The tubulin code and its role in controlling microtubule properties and functions. *Nat. Rev. Mol. Cell Biol.* 21, 307–326. doi: 10.1038/s41580-020-0214-3

John, E. R. (2001). A field theory of consciousness. *Conscious. Cognit.* 10, 184–213. doi: 10.1006/ccog.2001.0508

John, E. R. (2002). The neurophysics of consciousness. *Brain Res. Brain Res.* Rev. 39, 1–28. doi: 10.1016/S0165-0173(02)00142-X

Kalra, A., Benny, A., Travis, S., Zizzi, E., *et al.,* (2023). Electronic Energy Migration in Microtubules. *ACS Cent. Sci.* 9, 3, 352–361

Keppler, J., and Shani, I. (2020). Cosmopsychism and consciousness research: a fresh view on the causal mechanisms underlying phenomenal states. *Front. Psychol. 11:371.* doi: 10.3389/fpsyg.2020.00371

Keppler, J. (2021*)* Building Blocks for the Development of a Self-Consistent Electromagnetic Field Theory of Consciousness. *Front. Hum. Neurosci.* 15:723415. doi: 10.3389/fnhum.2021.723415

Kiefer, A., (2020). Psychophysical identity and free energy. *Journal of the Royal Society Interface, 17.*

Kim, J., (1990), *Explanatory exclusion, and the problem of mental causation,* in Information, Semantics and Epistemology, ed. E. Villanueva (Oxford: Basil Blackwell).

Kim, J., (1999), *Mind in a Physical World.* Cambridge, MA: MIT Press.

Kim, H., Hudetz, A., Lee, J., Mashour, G., Lee, U.; (2018). Estimating the Integrated Information Measure Phi from High-Density Electroencephalography during States of Consciousness in *Humans. Front Hum Neurosci.* ;12:42. doi: 10.3389/fnhum.2018.00042. PMID: 29503611; PMCID: PMC5821001

Kim, H., Lee, U., (2019) Criticality as a determinant of integrated information / in human brain networks. *Entropy* 21, 981

Kirk, R. (1974). Zombies vs materialists. *Proceedings of the Aristotelian Society (Supplementary Volume)* 48:135-52

Kitchener, P., Hales, C., (2022). What Neuroscientists Think, and Don't Think, About Consciousness. *Front. Hum. Neurosci.* 16:767612. doi: 10.3389/fnhum.2022.767612

Koch, C., (2006). Quantum mechanics in the brain. *Nature*: Vol 440

Koch, C., G. Tononi. (2008). Can machines be conscious? *Spectrum IEEE* 45: 55–59.

Koch, C. Massimini, M., Boly, M., Tononi, G., (2016). Neural correlates of consciousness: progress and problems. *Nat Rev Neurosci* **17,** 307–321. https://doi.org/10.1038/nrn.2016.22

Koch, C., (2022). Is Consciousness Fundamental? Interview with Robert Kuhn. *Closer to Truth*

https://www.youtube.com/watch?v=-yqr2spzBtM&t=80s
Lambert, N., Chen, Y.-N., Cheng, Y.-C., Li, C.-M., Chen, G.-Y., and Nori, F. (2013). Quantum biology. *Nat. Phys*. 9, 10–18. doi: 10.1038/nphys2474
Levine, J. (1983). Materialism and qualia: the explanatory gap. *Pacific Philosophical Quarterly*, 64: 354-361.
Libet, B. Gleason, C.A. Wright, E.W. Pearl, D.K (1983). Time of conscious intention to act in relation to onset of cerebral activity (readiness-potential): The unconscious initiation of a freely voluntary act, *Brain*, **106** (3) pp. 623-42 DOI: 10.1093/brain/106.3.623
Lycan, W., (1996). *Consciousness and Experience*. Cambridge, MA: MIT Press.
Lvovsky, A., Sanders, B. & Tittel, W. Optical quantum memory. *Nature Photon* **3**, 706–714 (2009). https://doi.org/10.1038/nphoton.2009.231
Ma, Y., Ma, YZ., Zhou, ZQ. *et al.* (2021). One-hour coherent optical storage in an atomic frequency comb memory. *Nat Commun* **12**, 2381. https://doi.org/10.1038/s41467-021-22706-y
Marshall, W., Albantakis, L., Tononi, G., Black-boxing and cause-effect power. (2018). *PLOS Computational Biology,* 14, e1006114. https://doi.org/10.1371/journal.pcbi.1006114.
McFadden, J. (2000). *Quantum Evolution*. London: HarperCollins.
McFadden, J. (2002). Synchronous firing and its influence on the brain's electromagnetic field: evidence for an electromagnetic theory of consciousness. *J. Conscious. Stud.* 9, 23–50.
McFadden, J. (2013). The CEMI field theory gestalt information and the meaning of meaning. *J. Conscious. Stud. 20, 152–182.*
McFadden, J. (2020). Integrating information in the brain's EM field: the cemi field theory of consciousness. *Neurosci. Conscious*. 2020, niaa016. doi: 10.1093/nc/niaa016
McFadden, J., (2023) Consciousness: Matter or EMF? *Front. Hum. Neurosci.* 16:1024934. doi: 10.3389/fnhum.2022.1024934
Mediano, P., Rosas, F., Luppi, A., Jensen, H., Seth, A., Barrett, A., et al. (2022a). Greater than the parts: a review of the information decomposition approach to causal emergence. arXiv:2111.06518 [q-bio.NC]
Mediano, P., Rosas, F., Bor, D., Seth, A., Barrett, A., (2022b). The strength of weak integrated information theory. *Trends in Cognitive Sciences, Vol. 26, No. 8 https://doi.org/10.1016/j.tics.2022.04.008*
Mørch, H., (2019). Is Consciousness Intrinsic? A Problem for the Integrated Information Theory. *Journal of Consciousness Studies* 26 (1-2):133-162(30).
Nagel, T. (1974). "What Is It Like to Be a Bat?". *The Philosophical Review*. **83** (4): 435–450. doi:10.2307/2183914
Oizumi, M., Albantakis, L., Tononi, G. (2014). From the phenomenology to the mechanisms of consciousness: Integrated information theory 3.0. *PLoS Comput. Biol.* 10, e1003588.
Parr, T., Da Costa, L., Friston, K.J. (2019). Markov blankets, information geometry and stochastic thermodynamics. *Philosophical Transactions of the Royal Society A Mathematical Physical and Engineering Sciences* 378(2164): 201905159.
Pearl, J., (1988). *Probabilistic Reasoning in Intelligent Systems: Networks of Plausible Inference*. Representation and Reasoning Series. San Mateo CA: Morgan Kaufmann. ISBN 0-934613-73-7.
Penrose, R. (1989). *The emperor's new mind.* Oxford. Oxford University Press
Penrose, R. (2019). Artificial Intelligence, Computation, Physical Law, and Consciousness. *Towards a Science of Consciousness. Conference proceedings:* https://www.youtube.com/watch?v=5pYmTMCvHM4
Penrose, R., (2021). Science and Roger Penrose. The Science of Consciousness (TSC), Webinar. https://www.youtube.com/watch?v=9YdNg2rEBW0
Pinotsis, D. A., and Miller, E. K. (2022). Beyond dimension reduction: stable electric fields emerge from and allow representational drift. *NeuroImage*, 119058. doi: 10.1016/j.neuroimage.2022.119058

Plato, (247), *Theaetetus. Sophist*, Plato in Twelve Volumes, Vol. 12 translated by Harold N. Fowler. Cambridge, MA, Harvard University Press; London, William Heinemann Ltd. 1921

Pockett, S. (2000). *The Nature of Consciousness: A Hypothesis*. Lincoln, NE: Writers Club Press.

Pockett, S. (2002). Difficulties with the electromagnetic field theory of consciousness. *J. Conscious. Stud.* 9, 51–56.

Pockett, S. (2012) The Electromagnetic Field Theory of Consciousness: A Testable Hypothesis about the Characteristics of Conscious as Opposed to Non-conscious Fields. *Journal of Consciousness Studies*, **19**, No. 11–12, pp. 191–223

Pokorný, J., Pokorný, J., Vrba, J., (2021). Generation of Electromagnetic Field by Microtubules. *Int J Mol Sci.;*22(15):8215. doi: 10.3390/ijms22158215. PMID: 34360980; PMCID: PMC8348406.

Popper, K., (1953). A note on Berkeley as precursor of Mach. *The British Journal for the Philosophy of Science*. **IV** (13): 13. doi:10.1093/bjps/IV.13.26

Popper, K. R. Eccles, J. C. (1977). *The Self and its Brain.* New York, NY: Springer

Primas, H., (2002), "Hidden determinism, probability, and time's arrow," in *Between Chance and Choice*, H. Atmanspacher and R.C. Bishop (eds.), Exeter: Imprint Academic, pp. 89–113.

Pylkkänen, P. (2007), *Mind, Matter and the Implicate Order*, Springer, Berlin.

Pylkkänen, P., Hiley, B., (2005) Can mind affect matter via active information? *Mind and Matter*, vol. 3, no. 2, pp. 7-26.

Pylkkänen P. (2019) Quantum Theory and the Place of Mind in the Causal Order of Things. In: de Barros J., Montemayor C. (eds) *Quanta and Mind*. Synthese Library (Studies in Epistemology, Logic, Methodology, and Philosophy of Science), vol 414. Springer, Cham. Doi: 10.1007/978-3-030-21908-6_14

Rahaman, F. (2021). Causal Structure of Spacetime. In *The General Theory of Relativity: A Mathematical Approach* (pp. 187-218). Cambridge: Cambridge University Press. doi:10.1017/9781108837996.009

Ramstead, M., Albarracin, M., Kiefer, A., Klein, B., Fields, C., Friston, K., Safron, A., (2023) The inner screen model of consciousness: applying the free energy principle directly to the study of conscious experience. *PsyArXiv*. doi: 10.31234/osf.io/6afs3.

Rao, R., Ballard, D., (1999). Predictive coding in the visual cortex: a functional interpretation of some extra-classical receptive-field effects. *Nat. Neurosci.* **2**, 79–87 (1999).

Ricciardi L. M.; Umezawa H. (2004) [1967]. Gordon G. G.; Pribram K. H.; Vitiello G. (eds.). *Brain physics and many-body problems*. Brain and Being. Amsterdam: John Benjamins Publ Co.: 255–266.

Rodriguez, E., George, N., Lachaux, J. P., Martinerie, J., Renault, B., & Varela, F. J. (1999). Perception's shadow: long-distance synchronization of human brain activity. *Nature*, *397*(6718), 430-433.

Rosas, F., Mediano, P., Gastpar, M., Jensen, H., (2019). Quantifying high-order interdependencies via multivariate extensions of the mutual information. Phys. Rev. E 100,

Rosas, E., Mediano, P., Jensen, J., Seth, A., Barrett, A., Carhart-Harris, R., Bor, D. (2020). Reconciling emergences: An information-theoretic approach to identify causal emergence in multivariate data*. PLoS Computational Biology* 16, e1008289.

Rosenthal, D. (1986). Two Concepts of Consciousness." In *Philosophical Studies* 49:329-59, 1986.

Rosenthal, D. (1997). *A theory of consciousness*. In N. Block, O. Flanagan, and G. Guzeldere, eds. The Nature of Consciousness. Cambridge, MA: MIT Press.

Ruffini G, Salvador R, Tadayon E, Sanchez-Todo R, Pascual-Leone A, Santarnecchi E (2020) Realistic modeling of mesoscopic ephaptic coupling in the human brain. PLoS Comput Biol 16(6): e1007923. https://doi.org/10.1371/journal.pcbi.1007923

Russell, B. (1913). On the notion of cause. *Proceedings of the Aristotelian Society* 13: 1–26.

Safron, A., (2020). An Integrated World Modeling Theory (IWMT) of Consciousness: Combining Integrated Information and Global Neuronal Workspace Theories with the Free Energy Principle and Active Inference Framework; Toward Solving the Hard Problem and Characterizing Agentic Causation*. Front. Artif. Intell.* 3:30. doi: 10.3389/frai.2020.00030

Safron, A., Çatal, O., & Tim, V. (2022). Generalized simultaneous localization and mapping (G-SLAM) as unification framework for natural and artificial intelligences: Towards reverse engineering the hippocampal/entorhinal system and principles of high-level cognition. *Frontiers in Systems Neuroscience*. https://doi.org/10.3389/fnsys.2022. 787659

Sahu, S., Ghosh, S., Ghosh, B., Aswani, K., Hirata, K., Fujita, D., et al. (2013a). Atomic water channel controlling remarkable properties of a single brain microtubule: correlating single protein to its supramolecular assembly. *Biosens. Bioelectron.* 47, 141–148. doi: 10.1016/j.bios.2013.02.050

Sahu, S., Ghosh, S., Fujita, D., & Bandyopadhyay, A. (2014). Live visualizations of single isolated tubulin protein self-assembly via tunnelling current. *Scientific Reports*, https://doi.org/10.1038/srep073034, 7303. https://www.nature.com/articles/srep07303

Sahu, S., Ghosh, S., Hirata, K., Fujita, D., and Bandyopadhyay, A. (2013b). Multilevel memory-switching properties of a single brain microtubule. *Appl. Phys. Lett.* 102:12370.

Saxena, K., Singh, P., Sahoo, P., Sahu, S., Ghosh, S., Ray, K., Fujita, D., & Bandyopadhyay, A. (2020). Fractal, scale free electromagnetic resonance of a single brain extracted microtubule. *Fractal and Fractional*, 4(11). https://doi:10:3390/fractalfract4020011

Sanfey, J., (1999), Footprints in knowledge: A third person phenomenology of consciousness. *Quantum Approaches to Consciousness. Conference proceedings*. Flagstaff. P51 [online] https://www.quantumconsciousness.org/sites/default/files/1999%20QM-1%20Quantum%20Approaches%20%20Program_0.pdf

Sanfey, J., (2003a), Time and observation. *Quantum Mind 2003: Spin-Mediated Consciousness Theory*. Tucson. Conference proceedings.; https://www.alice.id.tue.nl/references/2003-Quantum_Mind_Abstracts.pdf

Sanfey, J., (2003b), Reality, and those who perceive it. In *The nature of time: geometry, physics, and perception.* R. Buccheri, M. Saniga, W.M. Stuckey, (eds.) NATO Science Series. Dordrecht. Kluwer Academic Press.

Sanfey, J., (2005), The mind in physics. In *Endophysics, time, quantum and the subjective;* R. Buccheri. A.C. Elitzur and M. Saniga (eds.); pp 531–546. Singapore. World Scientific Publishing Co.

Sanfey, J., (2020). Time is the key for physical laws of consciousness. Conference presentation. *Towards a Science of Consciousness*, Tucson; 2020., https://www.youtube.com/watch?v=shf5vpdYUpo

Seagar, W,. (2006). The 'Intrinsic nature' argument for panpsychism: *Journal of Consciousness Studies*, 13, No. 10–11, 2006, pp. 129–45

Seagar, W,. (2019). Presence and Panpsychism. Towards a Science of Consciousness conference (TSC). Interlaken. https://www.tsc2019-interlaken.ch/program/plenary-seager/

Searle, J. (1992). *The rediscovery of the mind.* The MIT Press.

Seth, A., Barrett, A., Barnett, L., (2011) Causal density and integrated information as measures of conscious level. *Philos. Trans.* A 2011, 369, 3748–3767.

Seth, A., Suzuki, K.,Critchley, H. D. (2012). An interoceptive predictive coding model of conscious presence. *Frontiers in Psychology,* 2, 395.

Seth, K., (2014). A predictive processing theory of sensorimotor contingencies: Explaining the puzzle of perceptual presence and its absence in synesthesia. Cogn Neurosci; 5(2):97-118. doi: 10.1080/17588928.2013.877880. Epub 2014 Jan 21. PMID: 24446823; PMCID: PMC4037840.

Seth, A., (2015). Presence, objecthood, and the phenomenology of predictive perception. Cogn Neurosci; 6(2-3):111-7. doi: 10.1080/17588928.2015.1026888. Epub 2015 Apr 7. PMID: 25849361.

Seth, A., (2016). The hard problem of consciousness is a distraction from the real one. *Aeon Essays. Aeon.* Available online at: https://aeon.co/essays/the-hard-problem-of-consciousness-is-a-distraction-from-the-real-one

Silberstein, M., (1998). Emergence and the mind/body problem, *Journal of Consciousness Studies*, 5 (4), pp. 464–82.

Silberstein, M. & McGeever, J. (1999), The search for ontological emergence, *The Philosophical Quarterly*, 49, pp. 182–200.

Silberstein, M., (2001). Converging on Emergence: Consciousness, Causation and Explanation. *Journal of Consciousness Studies*, **8**, No. 9–10, pp. 61–98.

Singh, P., Saxena, K., Sahoo, P., Ghosh, S., and Bandyopadhyay, A. (2021a). Electrophysiology using coaxial atom probe array: live imaging reveals hidden circuits of a hippocampal neural network. *J. Neurophysiol.* 125, 2107–2116. doi: 10.1152/jn.00478.2020

Singh, P., Sahu, P., Ghosh, S., Bandyopadhyay, A., (2021b). Filaments and four ordered structures inside a neuron fire a thousand times faster than the membrane: theory and experiment. *Journal of Integrative Neuroscience.* 20(4), 777–790.https://doi.org/10.31083/j.jin2004082

Singh, P., Sahoo, P., Saxena, K., Manna, J., Ray, K., Ghosh, S., et al. (2021c). Cytoskeletal Filaments Deep Inside a Neuron Are Not Silent: They Regulate the Precise Timing of Nerve Spikes Using a Pair of Vortices. *Symmetry*; 13: 821.

Sperry, R. W. (1990). *Forebrain commissurotomy and conscious awareness;* in Brain Circuits and Functions of the Mind, ed. C. Trevarthen (New York, NY: Cambridge University Press), 371–388.

Stapp, H., (1993), *A quantum theory of the mind-brain interface*, in Mind, Matter, and Quantum Mechanics, Berlin: Springer, pp. 145–172.

Stapp, H., (2015,) *A quantum-mechanical theory of the mind-brain connection*, in Beyond Physicalism, E.F. Kelly et al. (eds.), Lanham: Rowman and Littlefield, pp. 157–193.

Stapp, H., (2023) Quantum Interactive Dualism: An Alternative to Materialism. (Accessed March 17th, 2023). https://www-physics.lbl.gov/~stapp/QID.pdf

Strawson, G., (2006). Realistic Materialism: Why Physicalism Entails Panpsychism, *Journal of Consciousness Studies*, 13(10–11): 3–31.

Stubenberg, L. Wishon, D,. (2023). Neutral Monism, *The Stanford Encyclopedia of Philosophy* (Spring 2023 Edition), Edward N. Zalta & Uri Nodelman (eds.), URL = <https://plato.stanford.edu/archives/spr2023/entries/neutral-monism/>.

Tegmark, M., (2000), Importance of quantum decoherence in brain processes, *Physical Review E* 61, 4194–4206.

Tegmark, M., (2016). Improved Measures of Integrated Information. *PLoS computational biology*, *12*(11), e1005123. https://doi.org/10.1371/journal.pcbi.1005123

Tononi, G. (2004). An information integration theory of consciousness. *BMC Neurosci.* **5**, 42. https://doi.org/10.1186/1471-2202-5-42

Tononi, G. (2008) Consciousness as integrated information: a provisional manifesto. *Biol. Bull.* **215**, 216–242.

Tononi, G. Koch, C, (2015), Consciousness: here, there and everywhere? *Phil. Trans. R. Soc.* http://doi.org/10.1098/rstb.2014.0167

Tononi, G. Boly, M. Massimini, M. Koch, C. (2016). Integrated information theory: from consciousness to its physical substrate. *Nature Reviews Neuroscience*. 17 (7): 450–461 DOI https://doi.org/10.1038/nrn.2016.44

Tononi, G., Koch, C., (2016). A reply to Barrett (2016). *Phil. Trans. R. Soc.* B 371: 20150452. http://dx.doi.org/10.1098/rstb.2015.0452

Tononi, G., Albantakis, L., Boly, M., Koch, C., (2022). Only what exists can cause: An intrinsic view of free will 10.48550/arXiv.2206.02069

Tuszynski, J. (2021), Measuring and modelling the electrical properties of microtubules. Bioelectrodynamic Webinars, BioEd 4, https://www.youtube.com/watch?v=WcWdNPP0gVU

Van Gulick, R., (2014/22) Consciousness, *The Stanford Encyclopedia of Philosophy* (Winter 2022 Edition), Edward N. Zalta & Uri Nodelman (eds.), https://plato.stanford.edu/archives/win2022/entries/consciousness.

Van Gulick, R., (2004) *Higher-Order Global States HOGS: An Alternative Higher-Order Model of Consciousness.* In R. Gennaro ed. Higher-Order Theories of Consciousness: An Anthology. Amsterdam and Philadelphia: John Benjamins, 2004.

Varela, F.J., Pachoud, B., & Roy, J. (2000). *The Specious Present: A Neurophenomenology of Time Consciousness.* In Naturalizing Phenomenology. Stanford University Press, Stanford Chapter 9, pp.266-329

Vitiello, G., (1995), Dissipation and memory capacity in the quantum brain model, *International Journal of Modern Physics B*, 9: 973–989.

Vitiello, G., (2001). *My Double Unveiled*, Amsterdam: Benjamins

Wu, W, (2018). The Neuroscience of Consciousness, *The Stanford Encyclopedia of Philosophy* (Winter 2018 Edition), Edward N. Zalta (ed.), https://plato.stanford.edu/archives/win2018/entries/consciousness-neuroscience/

Wheeler, J., (1990) Information, Physics, Quantum: The Search for Links. In: Zurek, W.H., Ed., *Complexity, Entropy, and the Physics of Information*, Addison-Wesley, Redwood City, 354-368

Wigner, E., (1961). *Remarks on the mind-body question*. In Good, Irving John (ed.). *The Scientist Speculates*. London: Heinemann. pp. 284–302. doi:10.1007/978-3-642-78374-6_20. ISBN 978-3-540-63372-3.

Young, A., Hunt, T., Ericson, M., (2022) The Slowest Shared Resonance: A Review of electromagnetic Field Oscillations Between Central and Peripheral Nervous Systems. *Front. Hum. Neurosci.* 15:796455.doi: 10.3389/fnhum.2021.796455

Zanardi. P., Tomka, M., Venuti, L., (2018). Quantum Integrated Information Theory: Comparison with Standard Presentation of IIT 3.0. *arXiv preprint arXiv:*1806.01421